# A Search for Transient, Monochromatic Light in a 6-deg Swath along the Galactic Plane

Geoffrey W. Marcy[1*] & Nathaniel K. Tellis[2]

[1] *Space Laser Awareness, 3388 Petaluma Hill Rd, Santa Rosa, CA, 95404, USA*
[2] *RocketCDL*

ABSTRACT

We searched the Milky Way Plane along a 6-deg swath for pulses of monochromatic light as faint as 15$^{th}$ mag (V band) using a wide-field telescope equipped with a prism. Pulses with duration less than 1 s that occur more often than once every 10 minutes would be detected, and pulses arriving less frequently would be detected with proportionally lower probability. A "difference-image" algorithm revealed 36 monochromatic sources. Subsequent assessment showed all were simply astrophysical objects emitting identifiable spectral lines.  No unexplainable monochromatic emission, pulsed or continuous, was detected. The detection threshold corresponds to a ~70 GW laser having a diffraction-limited 10-meter aperture located 1 kiloparsec away (depending on wavelength). Previous searches for laser emission from more than 5000 stars found none. Previous all-sky surveys at optical and radio wavelengths have revealed thousands of unexpected objects in the universe that exhibited extraordinary spectral emission, but none were technological. Hypotheses of our Milky Way Galaxy teeming with advanced life must be demoted.

## 1   INTRODUCTION

Theoretical models of extraterrestrial technologies involve interstellar communication by laser beams (e.g., Schwartz and Townes 1961, Bracewell 1973, Freitas 1980, Zuckerman 1985, Gillon 2014, Hippke, 2018, 2020, 2021ab; Gertz 2018, 2021, Gertz & Marcy 2022). Lasers offer many useful attributes including light-speed communication, energy-efficient narrow beams, high bit rates, and privacy.  The models predict that communication nodes, including relay stations, may be located throughout the Milky Way, creating a mesh of laser beams that constitute a galactic internet (Freitas 1980, Gillon 2014, Hippke 2020, 2021ab; Gertz 2018, 2021, Gillum 2022, Gertz & Marcy 2022). Indeed, lasers on satellites orbiting Earth are employed for the same reasons (Wang et al. 2023, Carver et al. 2023, Israel et al. 2023).

Searches for monochromatic light have previously been conducted toward more than 5000 stars of all types (O, B, A, F, G, K, and M) using spectra from the Keck 10-meter telescope and the Lick Observatory Automated Planet Finder, yielding no laser detections (Reines & Marcy 2002; Tellis & Marcy 2015, 2017; Marcy 2021; Marcy et al. 2021,2022; Tellis 2022 private communication; Zuckerman et al. 2023, 2024). Also, a search for optical laser emission was conducted within a thin, 2-deg wide swath along the Milky Way Plane (Marcy & Tellis 2023), finding no candidates. Searches were also done toward the solar gravitational lens focal points for nearby stars (Marcy et al. 2021). Powerfully, a search was done during planet transits of 218 exoplanet systems using NASA-Kepler photometry (Zuckerman et al. 2024).   No monochromatic light was found, neither sub-second pulses nor continuous emission.

Other searches for optical sub-second flashes were performed over a broad band of wavelengths (Horowitz et al. 2001, Werthimer et al. 2001, Wright et al. 2001; Howard et al. 2004; Stone et al. 2005; Howard et al. 2007, Hanna et al. 2009, Abeysekara et al. 2016, Villarroel et al. 2020, 2021, 2022a, 2022b, 2023, Maire et al. 2022, Acharyya et al. 2023, Solano et al. 2023).  Broadband optical searches can detect flashes from both extraterrestrial technology and from the mergers of black holes and neutron stars having a wide range of masses (cf. Wang 2022, Zhang 2020, Yan et al. 2019), motivated by Fast Radio Bursts and γ-ray bursts. The mergers are predicted to produce brief optical emission (Lyutikov 2017; Petroff et al. 2022; Zhang 2020; Yang et al. 2019).  Optical afterglows to GRBs are observed to last many hours (e.g., Sari & Piran 1999, Aasi et al. 2000), but shorter broadband optical pulses have not been detected.

Detecting unresolved broadband optical pulses must discriminate against false alarms such as glints of sunlight off satellites orbiting the Earth, strobe lights on distant aircraft, laser communication and LIDAR from orbiting satellites, and elementary particles hitting the sensor that mimic a star image (Corbett et al. 2021, Nir et al. 2021, Richmond et al. 2020). Villarroel et al. (2020, 2021, 2022a, 2022b) and Solano et al. (2023) have discovered transients, both in alignment and clumped, in pre-Sputnik images from the Palomar Sky Survey. Work is ongoing to identify these interesting candidates. Clearly, new SETI techniques are needed (e.g., Montebugnoli, Melis, Antonietti 2021)

To mitigate the challenges of broadband searches we search for monochromatic flashes, offering some discrimination from false alarms. Here we describe a search for pulses of monochromatic emission from a 6-degree wide swath along the plane of the Milky Way that is visible from the northern Hemisphere. This search also can detect long-lasting monochromatic emission, albeit with poorer detectability.

## 2 OBJECT-PRISM OBSERVATIONS OF THE MILKY WAY PLANE

Our previous spectroscopic survey of the plane of the Milky Way was confined to a swath 2 deg wide (Marcy & Tellis 2023), yielding detections only of known astrophysical objects with emission lines, notably Wolf-Rayet stars, Herbig Ae stars, compact planetary nebulae, and M dwarf flare stars. Here we enlarge that survey by observing with the same instrument the region 2 deg above and 2 deg below the previous survey along the Galactic Plane. This new survey is composed of 248 fields, each 2.1 x 3.1 deg, and each field was observed with 600 exposures of 1 sec duration. When combined with the previous survey, a total of 2587 sq. deg were observed spectroscopically, comprised of a 6 deg-wide swath along the Galactic Plane, shown in Figure 1. The swath includes the Galactic plane above Declination -32 deg, observable from our observatory at Taylor Mountain in northern California (see www.spacelaserawareness.org).

Figure 1 shows the RA and DEC of all the fields we observed, each field having of angular size 3.1 x 2.1 deg located along the Galactic Plane. We include the fields from the previous effort (Marcy & Tellis 2023). The swath extends from Galactic longitude -4 deg to +249 deg, observable from our Space Laser Awareness (SLA) facility in northern California. Also shown and labelled on Figure 1 is the 14 x 10 deg region we previously observed spectroscopically near the Galactic Centre (Marcy et al. 2022). Missing from this survey is the Galactic Plane south of DEC = -32 deg. The fields overlap to provide complete coverage, with the overlap regions observed twice.

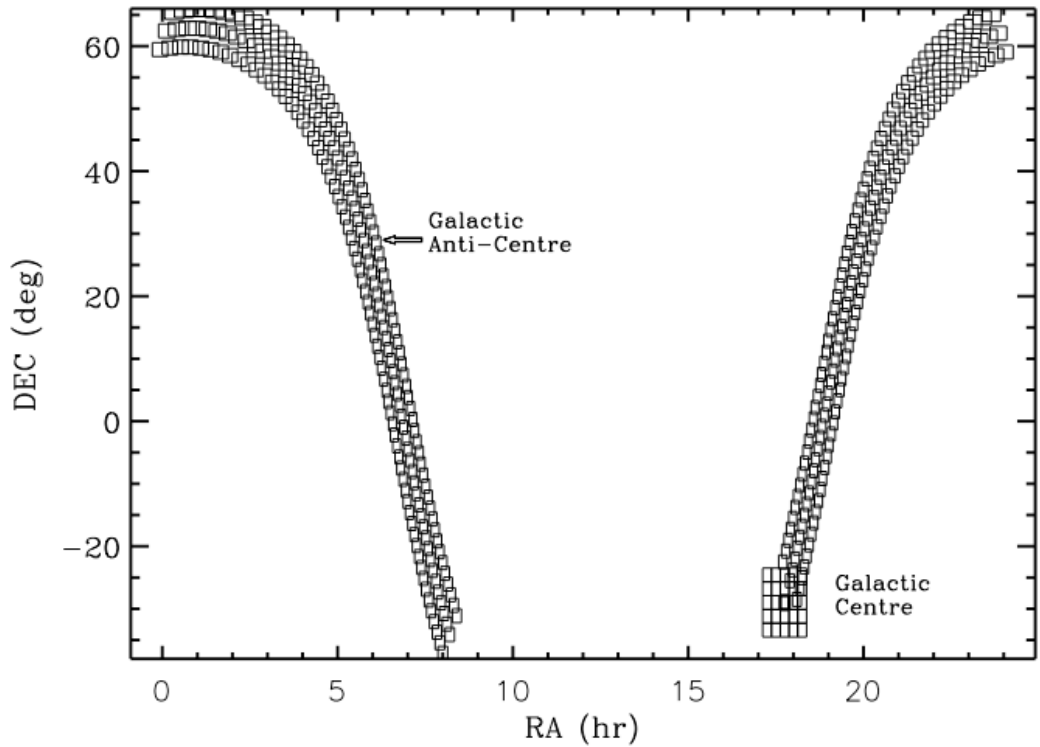

Figure 1. The 360 fields, each 3.2x2.1 deg, observed along the Galactic Plane in this objective prism survey. Each field was observed with 600 consecutive images of 1 sec exposure time, covering a total of 2750 sq. deg. Each exposure contained spectra of hundreds of stars and of any sources between the stars, as shown in Figure 2.

We used the objective-prism Schmidt telescope with an aperture of 0.28 m described in Marcy & Tellis (2023). The telescope has a focal ratio of f/2.2 and focal length of 0.62 m. Focussing is performed nightly by pistoning the primary mirror, and the image quality over the entire field was optimized by using an Octopi-Astro unit to adjust the piston and tip/tilt of the camera, resulting in a PSF having a FWHM = 4 pixels (5 arcsec) in best seeing. A 7-degree angle wedge prism was placed over the front of the telescope, to produce spectra of low resolution, R~100, of every point within the field of view of 2.1 x 3.1 deg. The typical PSF has FWHM = 4 to 6 pixels, which is a key design feature. The over-sampling allows true point sources to be distinguished unambiguously from cosmic rays and natural radioactivity that only pollute a few adjacent pixels. This over-sampled PSF allows a monochromatic pulse to be unambiguously detected in just one exposure.

The QHY600 CMOS camera at the focal plane contains 9600 x 6420 pixels, each 3.7 microns square that subtend 1.3 arcsec. The quantum efficiency is ~80% between 500 – 800 nm and more than 10% between 400 nm and 900 nm. We operated the QHY600 CMOS sensor with a "Gain"=57 and "offset"=12 that yields a system gain of 0.33 e-/ADU. Hereafter we use "count" for "ADU" and often convert immediately to the number of photons. This CMOS system has a read noise of ~2 e- and a dark noise of ~0.02 e- $px^{-1}$ $s^{-1}$. We operated the camera at a temperature of -10C and full well of 22000 e-, and it offers linearity better than 1% (Alarcon et al. 2023). About 2% of the individual pixels are sporadically high or low by ~30 e-, due to readout electronics, a standard "salt and pepper noise" that is easily removed by median smoothing if needed. We operate with a frame rate of 1 fps with fiber-optic data transfer between the camera and PC, that imposes dead time less than 0.01 s between frames (Gill et al. 2022, Betoule et al. 2022, Alcaron et al. 2023). We read the 60 Megapixel images with 2 bytes per pixel. We observe with no filter to capture spectra from 380 to 950 nm. The sky contributes ~40 photons per pixel per sec, depending on lunar phase, with most sky light coming from Santa Rosa city lights, causing a Poisson noise of ~7 photons (RMS) in each pixel.

Reflections of sunlight off satellites (Corbett et al. 2021, Nir et al. 2021) exhibit the full optical solar spectrum and they move across the field of view, allowing immediate rejection as not monochromatic. The 1-sec exposures provide a "movie" of all moving objects, including the angular velocity vector of the satellites as faint as 15[th] magnitude yielding an orbital ephemeris. Similarly, the motion of aircraft is temporally resolved in successive 1-sec exposures. False positives could come from laser pulses emitted by a satellite, especially pulses shorter than 0.1 s from satellites in geosynchronous orbit that may avoid showing elongation of the image, thus masquerading as an interstellar laser pulses. Laser pulses from satellites remain as prospective false positives. As we describe in the results section, we did not detect any such laser candidates at all, so confusion from Earth-bound satellites did not arise.

The optical design used here is similar to the Harvard objective prism telescopes used over 100 years ago (e.g., Pickering 1912; Fleming et al. 1917, Cannon & Pickering 1922). That survey of 225,000 stars would have detected persistent laser emission, if brighter than 10[th] mag. But the only unidentified emission lines they detected (as far as we know) were the atomic lines in Wolf-Rayet stars and the "nebulium" lines, originally discovered by Huggins and Miller (1864) and named by M. Huggins (1898). Of course, those were later shown to arise from forbidden transitions in ionized oxygen atoms at 372.7 nm and 500.7 nm. See Section 6 for implications of this non-detection.

Figure 2 shows a typical image from a 1-second exposure obtained with the objective prism telescope system and the QHY600 CMOS camera. The image exhibits hundreds of stellar spectra oriented vertically, spanning wavelengths 380 – 950 nm spread over 1200 pixels, with long wavelengths upward and North up. In our 1 sec exposures, the faintest stars have Vmag=14 with signal-to-noise ratios of ~5 per pixel. Laser emission would appear either in between the stellar spectra or directly superimposed on stellar spectra. The stellar absorption lines provide the local wavelength scale.

This particular image in Figure 2 contains the planetary nebula NGC6803, found automatically by our algorithm (Section 3) in its search for monochromatic lines, in this case the Balmer and [OIII] emission lines. NGC6803 has Bmag=13.0 mag, Vmag=11.6, and Rmag=13.2, representing the secure detection threshold of our instrumentation and analysis (see Section 5) for persistent (not pulsed) sources. Indeed, this image in Figure 2 is representative of the 1-second exposures along the Galactic plane in the density of stars and the appearance of astrophysical objects that emit emission lines, showing that non-astrophysical monochromatic emission would also be detected.

Many of the images revealed spectra of compact planetary nebulae, active M dwarfs with Balmer emission, Herbig Ae and Be stars, and Wolf-Rayet stars. Monochromatic and spatially unresolved point sources appear as a two-dimensional "dot" with a PSF shape. Sub-second monochromatic pulses would appear in only one image as a PSF-shape "dot" within a sequence of images, detectable by difference-imaging. The detection threshold in one 1-sec exposures is near magnitude 15 (see Section 5), and stars brighter than Vmag = 2.5 saturate the sensor with >56000 photons/pixel.

The individual spectra in the images can be wavelength-calibrated from stellar absorption lines and flux-calibrated using Vega as a reference (Marcy & Tellis 2023). The spectral resolution is set by the PSF that has FWHM ~5 pixels for RA = 2hr to 8 hr and FWHM ~ 4 pixels for RA = 18hr to 2hr. The latter set benefitted from the introduction of an "Octopi" tip-tilt device offering an optimized "back-focus" spacing for the CMOS sensor relative to the lens group. The resulting spectral resolution is modest, 2 to 10 nm from 370 to 950 nm, respectively, which is adequate to identify the usual astrophysical emission lines.

Sunlight glints off satellites (or rocket boosters) that are orbiting Earth are easily identified both by their solar spectrum and by their obvious motion during the sequence of 1 second exposures, rejecting them as extraterrestrial laser pulses. The spectroscopic information allows instant discrimination of both astrophysical and terrestrial sources from monochromatic extraterrestrial sources. Laser pulses from human-made satellites could be indistinguishable from extraterrestrial laser pulses, if the pulse duration is sufficiently short.

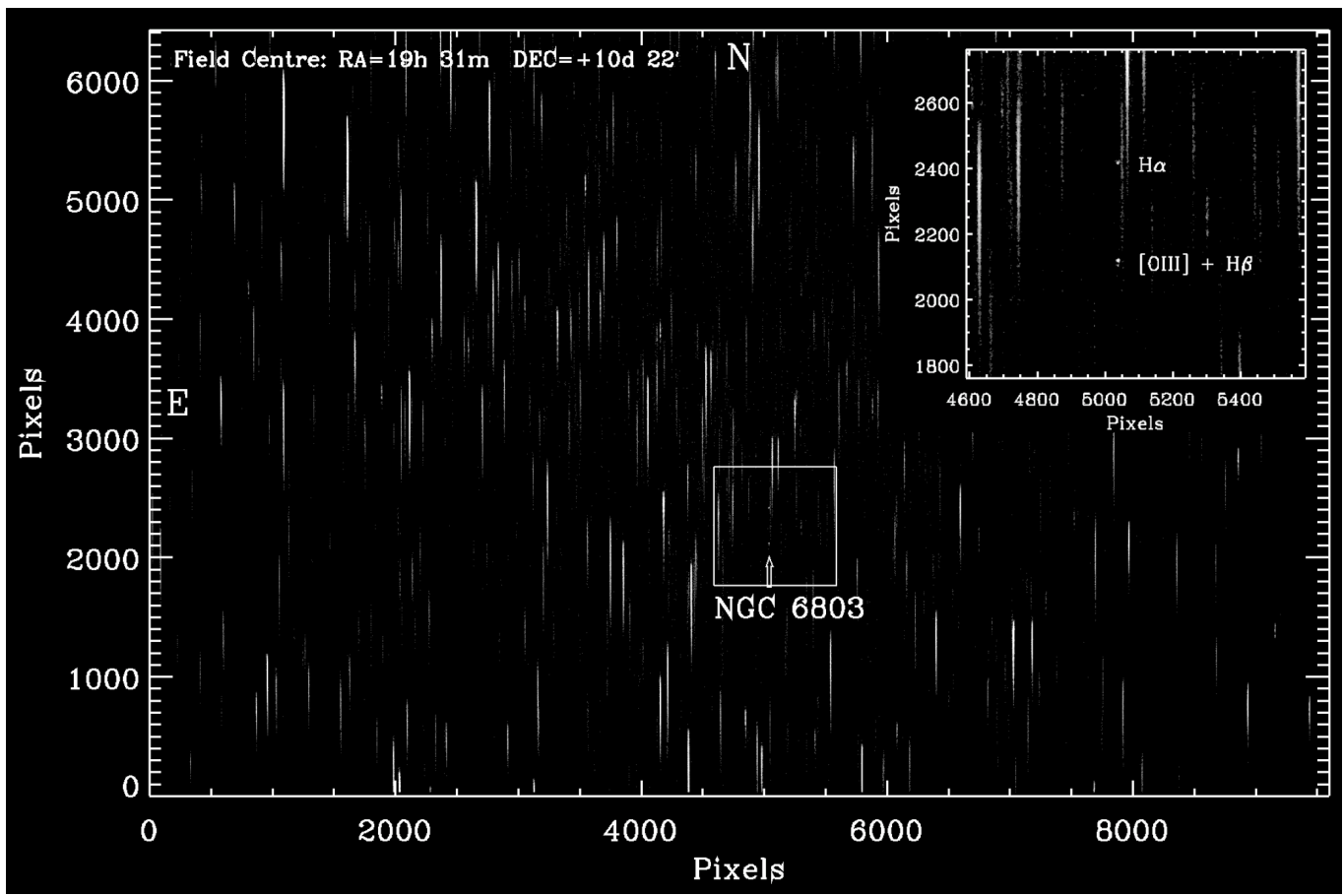

Figure 2. A representative 1-sec exposure with the objective prism system. Vertical lines are stellar spectra, brighter than V=14 mag, spanning wavelengths 380 to 950 nm, with long wavelengths upward. The field of view is 3.1 x 2.1 deg, spanning 9600 x 6400 pixels, each subtending 1.3 arcsec on the sky. A compact planetary nebula, NGC6803 (V=11.6 mag), was "discovered" in this 1-sec exposure by the automated analysis (Section 3). The upper-right shows a zoom of NGC 6803.

3   THE DIFFERENCE-IMAGE ALGORITHM

A simple difference-image technique is used to detect monochromatic emission from spatially unresolved sources, revealed as a "dot" having a nearly PSF shape. The algorithm proceeds by computing the difference between each image and the average of six "bookend" images, the three images prior and three after. The resulting difference image has the stars removed, leaving only sources that appeared, or brightened, during the 1-sec exposure. The algorithm suppresses residuals along stellar spectra caused by atmospheric scintillation by executing a 50-pixel boxcar smoothing along the direction of dispersion and subtracting it from the original difference image. The residuals are typically tens of photons in each pixel, due to Poisson noise of the sky and stellar photons, electronic noise of the CMOS sensor, and seeing variations.

Newly appearing emission lines located coincident with any stellar spectrum, or located in between stellar spectra, stand out in the difference images. This algorithm has several free threshold parameters that establish the minimum signal-to-noise ratio of the emission, the minimum chi-squared value of the profile of the candidate dot to the actual 2D PSF, and a minimum number of photons. These parameters were carefully set to make the usual trade-off between minimizing the number of false positives against a maximum sensitivity to monochromatic emission. The trade was set to yield one laser "candidate" per 2000 images. These threshold parameters were set by running hundreds of trials created by digitally superposing synthetic monochromatic emission into actual images. Figure 3 shows a zoom of seven consecutive actual images. A synthetic monochromatic emission pulse (a dot) was injected into the 4$^{th}$ image prior to executing the difference-image algorithm. This emission pulse is actually the H$\alpha$ emission we observed in Campbell's Star, an unresolved planetary nebula (See Section 4), and it contains the actual asymmetries and non-Gaussianity of the PSF more accurately than a synthetic 2D Gaussian.

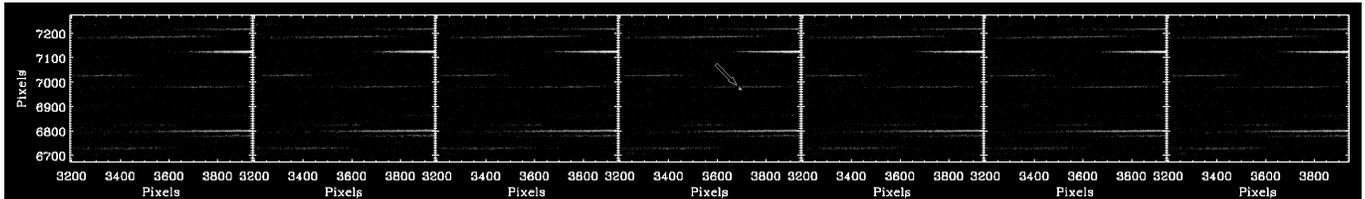

*Figure 3. A representative set of 7 consecutive exposures, each 1-sec, zoomed to 600x600 pixels, that includes a mock laser pulse that is PSF-shaped and placed at a random location in the image. The difference-image algorithm blindly found this mock laser pulse (arrow). The horizonal lines are actual spectra of stars from the objective prism telescope. This laser pulse contains 284 photons, corresponding to 15$^{th}$ magnitude (during 1 s), injected into an observed image.*

For each exposure, the algorithm determines the FWHM of the spatial profile of stellar spectra, commonly 4 to 6 pixels, caused by seeing and optical aberrations in the prism. Elementary particles that hit the CMOS sensor are ignored, as they are identified by containing electrons in only a few pixels, unlike the PSF.

The difference-image algorithm will fail to detect sub-second pulses having a cadence more frequent than 1 pulse per second, as both the target image and reference image will contain the emission. However, if the pulse intensities vary during 7 s, or if the seeing changes during 7 s, the algorithm may still detect it. Monochromatic light that is intrinsically constant in time is usually detected in the observations by the difference-image algorithm due to seeing changes during the sequence of 600 1-sec exposures that occasionally cause intensity changes of ~10%. Indeed, most of the monochromatic candidates are of that nature.

## 4 DETECTED MONOCHROMATIC CANDIDATES

The difference-image algorithm was executed on all 248 fields above and below the Milky Way Plane, including 600 exposures of 1 sec duration at each field (Figures 1 and 2). The automated difference algorithm identified 36 candidates of monochromatic emission. Extraction of the full spectra, covering wavelengths 380 to 950 nm, was performed for each candidate, allowing an analysis of each spectrum. The spectrum extraction process was simple, involving a subtraction of the local sky brightness (median) and a sum of 10 pixels perpendicular to dispersion at each location along the length of the spectrum of ~1500 pixels. Optimal extraction is not necessary as spectrophotometric accuracy is not relevant to identifying the nature of the object. The goal was to detect any associated continuum flux and any other emission lines to help identify the physical nature of the emission source.

Any stellar continuum in the extracted spectrum offers an opportunity to identify the stellar absorption lines thereby setting the wavelength scale, especially the zero-point, that determines the wavelength of the emission line found blindly by the difference-image analysis. If that wavelength is one commonly found in astrophysical sources, e.g., H-alpha, [OIII], or known lines in Wolf-Rayet stars, the line-emitting region is likely ionized hot gas found commonly among astrophysical objects, such as Herbig Ae or Be stars, planetary nebulae, Wolf-Rayet stars, or magnetically active M dwarf flare stars.

Similarly, any other emission lines that exhibit a known pattern offer a wavelength tag for all of them, thereby setting the entire wavelength scale. Common patterns of emission lines include Balmer lines, [OIII] lines, and spectral lines from different classes of Wolf-Rayet stars. Such patterns of lines allow an identification of the type of astrophysical object, ruling out non-standard explanations such as a technological origin.

Any monochromatic emission that has neither an associated continuum nor any other emission lines would survive as a possible non-astrophysical source such as a laser. Similarly, any unrecognizable pattern of emission lines would survive as such a candidate. Here we present representative cases of all of the emission lines identified by the automated algorithm. None survived as plausibly non-astrophysical.

Figures 4 and 5 show a common type of monochromatic emission found in our survey of the Galactic Plane by the automated difference-image analysis. Figure 4 shows a raw image from a 1-second exposure that exhibits four emission lines, with increasing wavelength to the right and north up. The pattern immediately resembles Hα, Hβ, and the barely resolved [OIII] lines at 495.9 and 500.7 nm. Hα (656.3nm) and [OIII] (500.7 and 495.9 nm) are always separated by 300 pixels, making that pattern immediately recognizable. Using the relative wavelength scale derived from the spectra of A-type stars, a wavelength zero-point immediately presents itself that confirms those astrophysical wavelengths, shown in Figure 5. The spectrum resembles the spectra of compact planetary nebulae. Astrometry shows this source is the compact planetary nebula, NGC6803, having V = 11.6 mag. It resides at a distance of 1.9 kpc and has angular size of 6 arcsec.

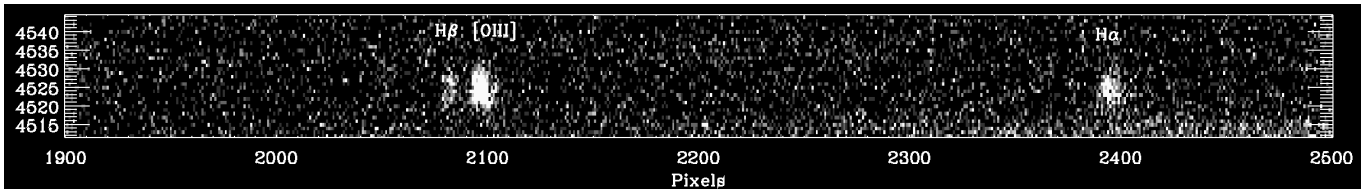

*Figure 4. A zoom of a 1 second exposure of the compact planetary nebula, NGC6803, Vmag=11.1, showing the four common emission lines spaced as expected for H-α, the [OIII] lines at 495.9 and 500.7 nm, and H-β. The extracted 1-D spectrum is shown in Figure 5. A 1-sec exposure reveals the pattern of astrophysical emission lines, ruling out non-astrophysical explanations including lasers*

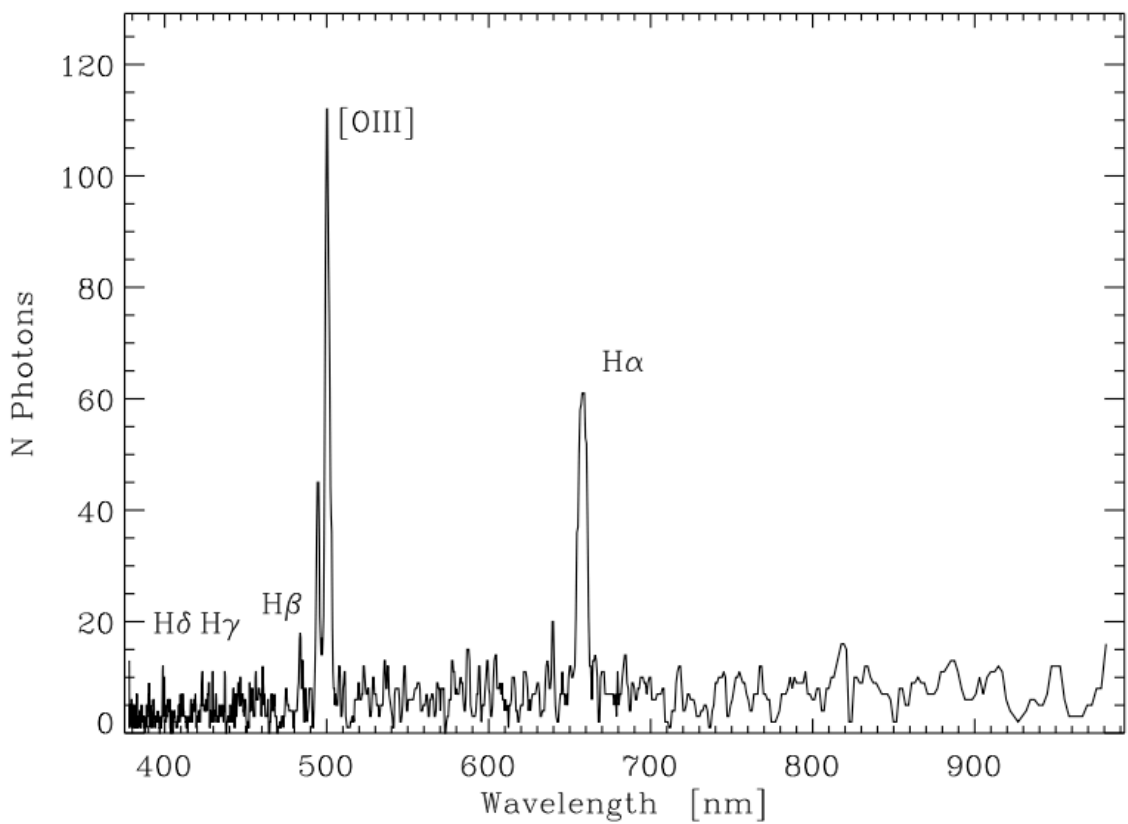

*Figure 5. The extracted spectrum of the raw image shown in Fig.4 containing a candidate monochromatic emission source. Four emission lines are prominent corresponding to Hα, Hβ, and the [OIII] lines at 495.9 and 500.7 nm, indicating the source is merely a common planetary nebula, indeed NGC6803 at V = 11.6 mag. This discounts any non-astrophysical explanation. Constant monochromatic flux is detected robustly by the difference-image algorithm, and fainter pulses will also be detected.*

This candidate monochromatic emission in Figure 5 offers two messages. The emission line was identified in 35 out of 600 exposures as "transient" monochromatic emission despite having constant intensity. The "detections" were due to seeing variations. The difference algorithm combined with seeing variations among the 600 exposures efficiently detects monochromatic emission from V=11.6 mag sources despite being long-lived, i.e., having a duration of many seconds, minutes, or indefinitely. This astrophysical emission-line source was detected by the difference-image algorithm due to sub-second seeing changes that momentarily raised the intensity of an emission line by 5 to 10% but that actually is constant in intensity.

Thus, non-astrophysical emission lines that are long-lived will be detected efficiently, among the 600 exposures. Individual pulses of ~10% as bright will also be detected, indicating a detection threshold near Vmag ~ 15 (see Section 5).

The second message is that this planetary nebula at Vmag=11.6 was easily detected in 35 exposures by this objective prism system that is optimized for monochromatic sources that *change* intensity. Of course, the emission-line intensity changed by only a fraction of the Vmag=11.6 flux due to seeing variations, triggering the difference-image algorithm. Thus while constant monochromatic emission of 12th magnitude is clearly detectable due to 10% seeing variations, a single, sub-second pulse has a detection threshold near Vmag~15, which we quantify carefully in Section 5.

A similar example is the candidate monochromatic emission shown in Figure 6, which upon analysis of the spectrum and astrometry is the planetary nebula, "PN Hb 12" with V = 11.6 mag (Acker et al. 1992). The first two Balmer lines are prominent as are the [OIII] lines (495.9 and 500.7 nm). Again, the message is that monochromatic emission of 12th magnitude, whether steady or pulsed with a duration under 1 second, would be detected in this survey. All emission-line objects brighter than Vmag=12 show up in this Galactic Plane survey, demonstrated by the blind detection of well-known planetary nebulae.

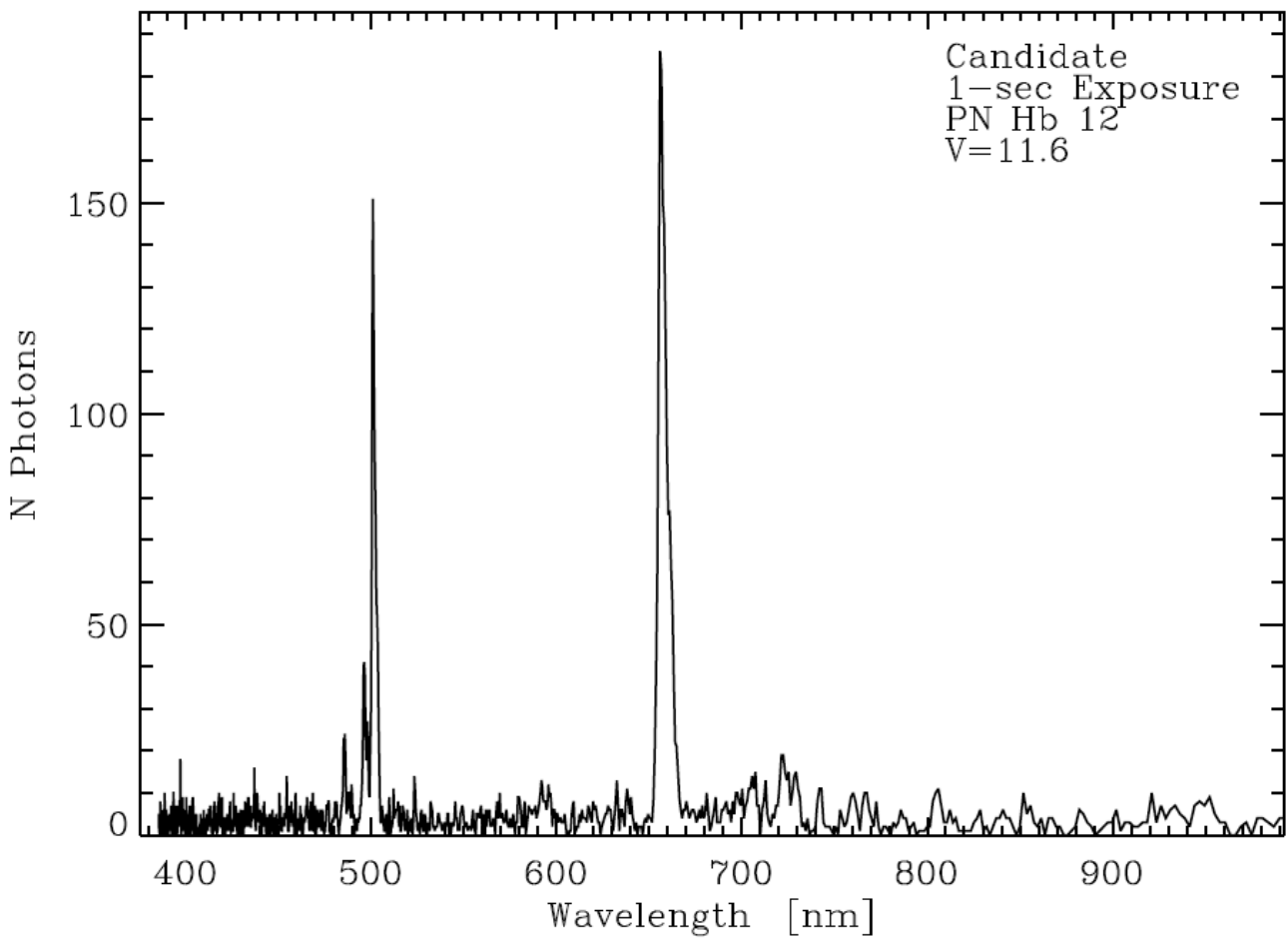

*Figure 6. A candidate monochromatic emission found automatically by the difference-image system. The spectrum is clearly that of a planetary nebula, and astrometry shows it to be the known planetary nebula, PN Hb 12 with Vmag=11.6, near the secure detection threshold.*

Figures 7 and 8 exhibit another representative detection from the difference-image analysis corresponding to the bright end of detectability. The difference-image algorithm "discovered" the monochromatic Hα emission in the Be star, γ Cas. Figure 7 shows a sequence of 7 consecutive images zoomed on γ Cas exhibiting its strong Hα emission that was identified blindly by the difference-image algorithm. The emission is intrinsically constant on this time scale, but seeing variations cause the detected intensity to vary by ~10%, allowing the image-difference algorithm to detect it. Figure 7 shows the monochromatic emission has a distorted shape, which is due to the poor PSF in the far corner of the 2.1 x 3.1 deg field of view. Indeed, such a wide field of view presents optical challenges to the RASA11 telescope, despite tip-tilt adjustment of the camera, with an *Octopi*.

Figure 8 shows the extracted spectrum from Figure 7, showing two Balmer lines and three telluric absorption bands, all of which provide the approximate wavelength scale shown in the plot.  At V = 2.5 mag, these objective prism spectra of γ Cas are nearly, but not quite, saturated by the QHY600 CMOS camera, thus allowing laser detection at such levels of brightness.

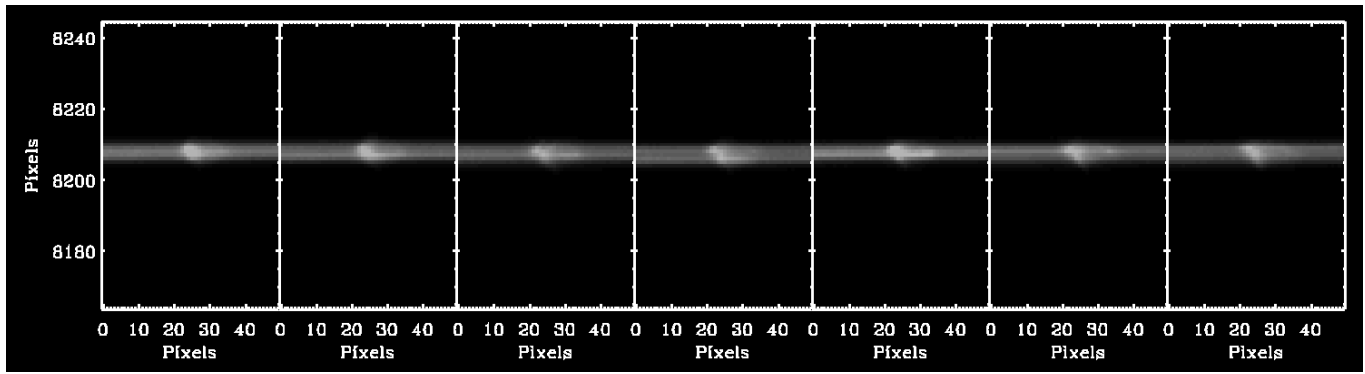

*Figure 7. A sequence of seven consecutive raw images (1 sec exposures) zoomed on a candidate monochromatic emission found blindly by difference-image analysis.  The star is γ Cas, a known Be star. It shows a stellar continuum and emission line at Hα, based on a wavelength scale established from the full stellar spectrum (Fig 8).  The emission line exhibits a distorted PSF due to its location in the far corner of the image, 1 deg north, and 1.5 deg east of the center of the field.*

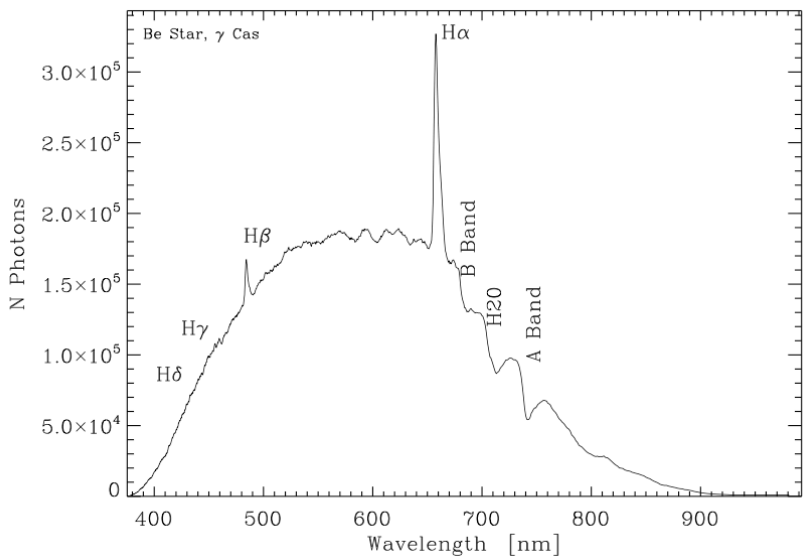

*Figure 8. The extracted spectrum of the monochromatic emission found by the difference-image analysis. Two Balmer lines in emission are apparent, along with three telluric absorption lines, all of which set the wavelength scale.  Astrometry shows this is the well-known Be star, γ Cas.  Seeing variations caused this to be detected as a transient emission line, despite its intrinsic constant intensity.*

Figure 9 shows another representative among the monochromatic candidates that emerged blindly from the difference-image analysis in this Galactic Plane spectroscopic survey. The strong emission lines at both 470 nm and 650 nm triggered the difference-image algorithm due to seeing variations, despite their intrinsically constant intensity. The spectrum (Figure 9) contains a suite of emission lines immediately identifiable as those from a WN6 Wolf-Rayet star, allowing the full wavelength scale to be interpolated and used in the plot. Astrometry of the raw image shows this to be HD192163, indeed of spectral type WN6 (SIMBAD) having Vmag = 7.5.

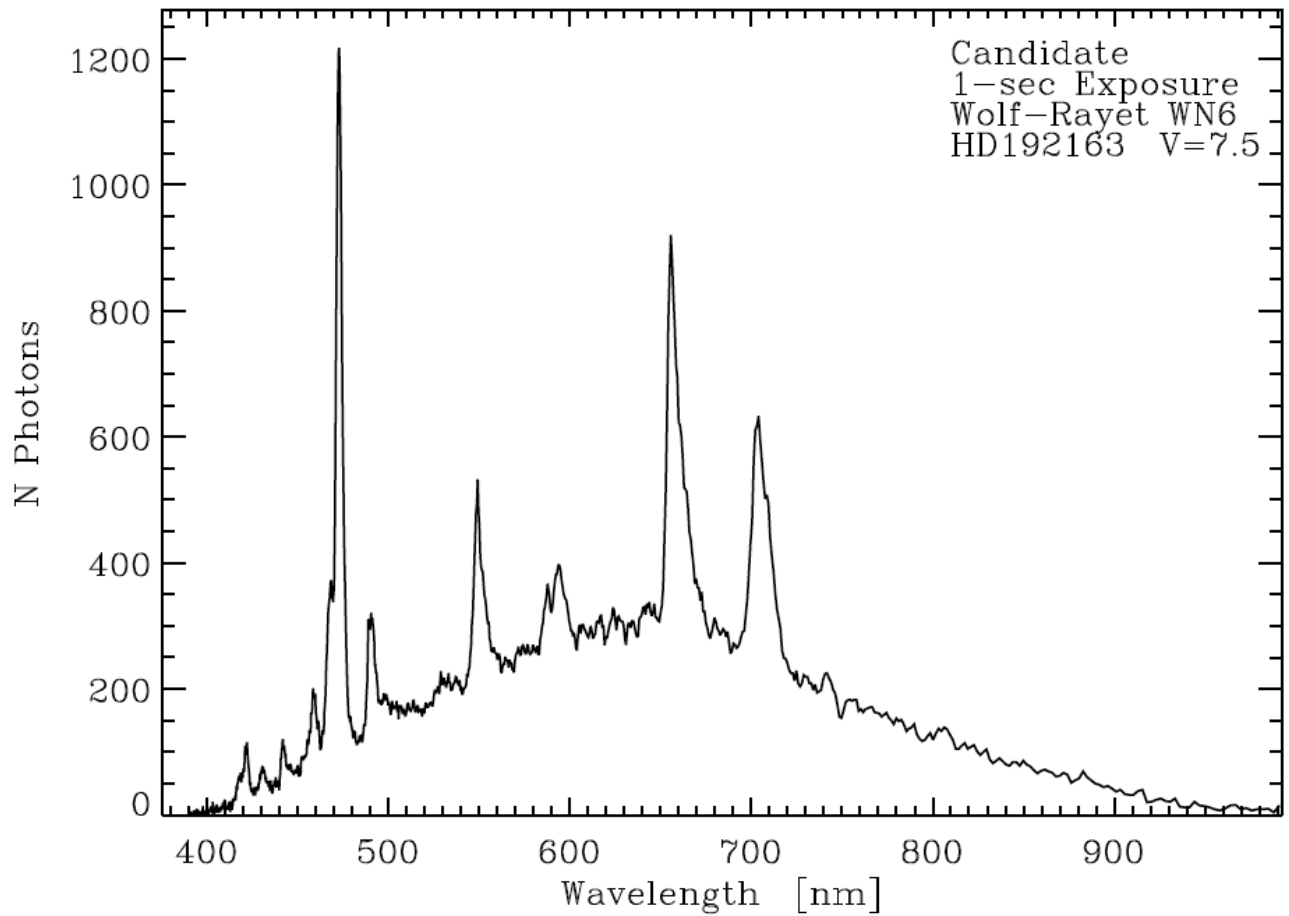

*Figure 9. A candidate source of monochromatic emission identified by the difference-image algorithm. The spectrum and its coordinates reveal it to be a well known Wolf-Rayet star of type WN6, namely HD192163, with Vmag=7.5. The survey "found" all 11 of the known Wolf-Rayet stars brighter than Vmag = 11.*

Figure 10 shows another candidate emission-line object discovered in the Galactic Plane survey. The raw images exhibited one strong emission line and six much weaker emission lines in each 1-second exposure. The emission lines appear in all 600 of the 1-sec exposures, with constant intensity within ~10%, consistent with seeing variations. Astrometry was performed by using two reference stars, Phi Cygni and 9 Cygni, yielding an RA scale of 0.0938 sec per pixel and a DEC scale of 1.266 arcsec per pixel. These give the coordinates of the emission-line candidate, RA = 19h 34m 45.2s +30d 30' 59" (2000), the location of "Campbell's Star", a dense planetary nebula with a Wolf-Rayet (WC9) central star having a total brightness, V=10.4 mag. Campbell's star offers an example of a spectrum sufficiently rare that the suite of emission lines could be naively misinterpreted as non-astrophysical (discussed further in Section 6).

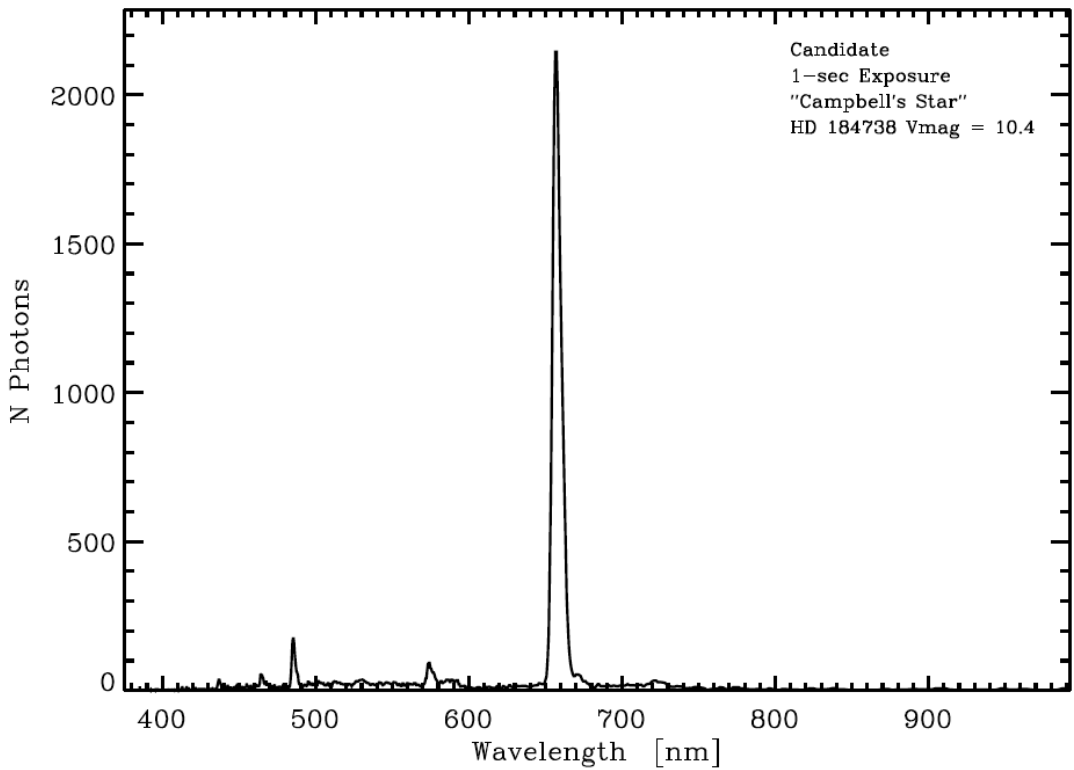

*Figure 10. A 1-sec exposure revealing a candidate emission-line source found by the image-difference analysis. Astrometry shows this is a previously discovered planetary nebula with a Wolf-Rayet central star known as "Campbell's Star", HD 184738, with Vmag=10.4. This case supports the contention that steady laser emission, and laser pulses with cadence faster than 1 pulse/sec, would have been discovered in past surveys as faint as Vmag=12.*

The automated difference-image code identified some candidate "emission-lines" that were actually wavelength regions of a stellar continuum where the flux is naturally high. A momentary improvement in the seeing caused the high continuum flux to appear even higher. One example is shown in Figure 11 with a zoom in Figure 12. This is the spectrum of an M dwarf having R = 9 mag at RA = 18h 27m  Dec=-21d 08' (2 deg below the Galactic Plane at longitude 5 deg), where the stellar continuum apparently brightened near 710 nm during one of the 1-sec exposures. We measured a 7.4% increase in the apparent continuum intensity at 710 nm, having a width of ~30 pixels, equal to that of the feature in the normal stellar continuum flux feature. This width of 30 pixels is greater than the spectral resolution of ~4 pixels, indicating that the entire stellar continuum brightened rather than an appearance of an emission line lasting 1 second.

More generally, spectra of cool stars, Teff < 4000, exhibit peaks and valleys that could be naively interpreted as due to a forest of monochromatic emission lines. However, comparison with catalogs of M dwarf spectra immediately allows the correct interpretation as an M dwarf or M-type giant star of some specific spectral type and metallicity (e.g. Leggett et al. 2000). Particularly identifiable for M3 to M8 dwarfs is the peak in the spectral energy distribution at 710nm and the double-hump peak at 770 nm. The peaks are well known to be caused by absorption outside those regions by TiO, CaH, VO, CaOH and other simple molecules.

Both scintillation of the star's light and a greater confinement of the stellar continuum due to improved seeing can produce this brief enhancement in stellar continuum flux. As just discussed, M dwarfs exhibit a natural peak in their continuous spectrum at ~710 nm due to absorption bands from CaH and TiO shortward of 710 nm and TiO longward of 710 nm (cf. (Leggett et al. 2000, Koizumi et al. 2021). The resulting peak in the measured flux at 710 nm during 1 second can be momentarily enhanced by a brief improvement in seeing, as shown in Figure 11 and 12.

The difference-image code identified candidate emission lines in some A- and F-type stars at the wavelengths of their Balmer absorption lines. The algorithm was fooled when the seeing was momentarily worse, which smeared flux from the stellar continuum into the core of the absorption line, raising the flux in the core. Examination of those spectra by eye revealed this effect as not real.

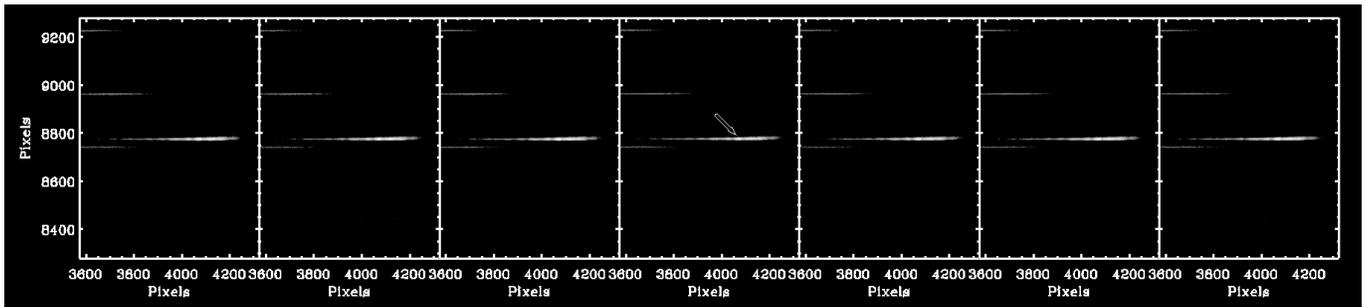

*Figure 11. Seven consecutive exposures of an M dwarf spectrum showing a 7.4% rise in its continuum flux at a small range of wavelengths.  The rise is due to a momentary improvement in seeing that causes a spurious increase in photon flux during one second. This is a common false positive.*

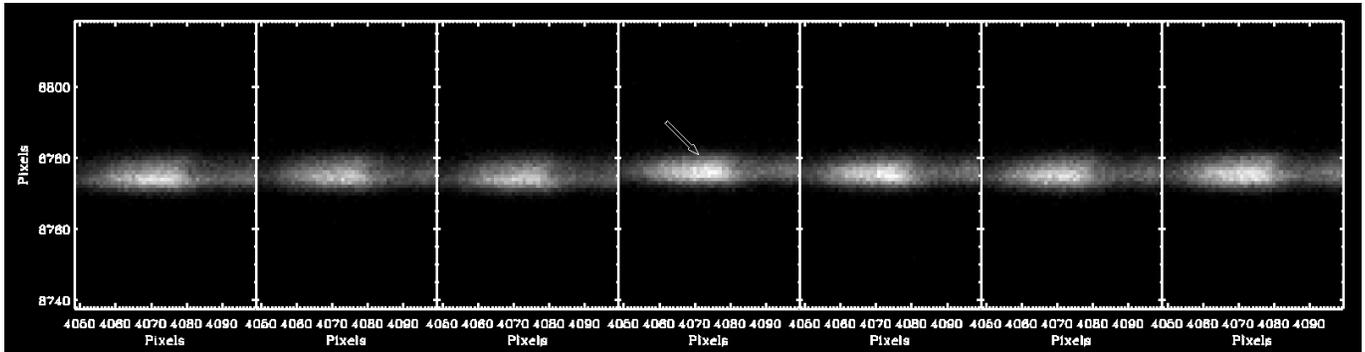

*Figure 12.  A zoom of Figure 11, showing the same 7 consecutive exposures of an M dwarf spectrum.  A brief improvement in seeing caused a false indication of a brief increase in the intensity of 7.4% of the continuum flux from the star. Such apparent brightenings of ~10% in intrinsically choppy spectra are a common occurrence, and dismissed as due to seeing.*

In our survey, aircraft, satellites, and meteors are commonly detected by the image-difference algorithm, with one example shown in Figure 13.  The brightness, angular velocity, and direction of motion are immediately apparent in the succession of images.  Often multiple lights are associated with the object, each light yielding a real-time spectrum including a spectrophotometric measurement. Monochromatic sources such as lasers are immediately apparent.  The displacement from frame to frame yields the two-dimensional angular velocity and direction.  Unidentified sources brighter than 12[th] mag show up easily.

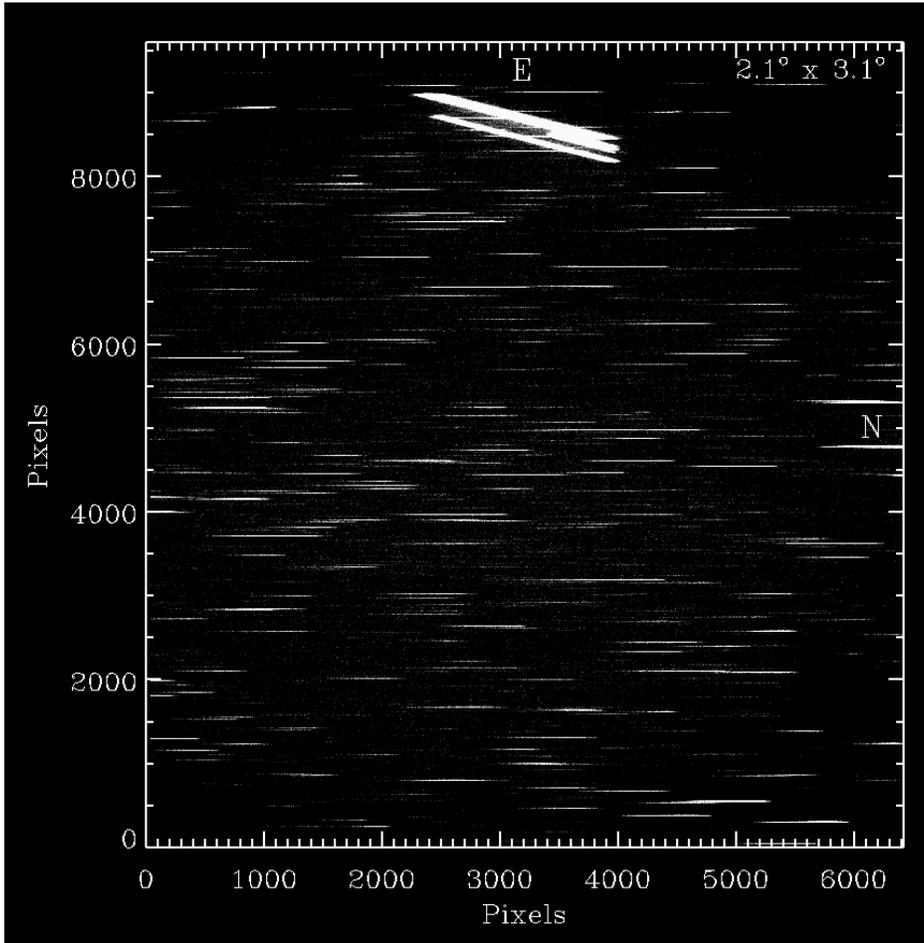

*Figure 13. A representative aircraft found blindly by the difference-image algorithm in a 1-sec exposure. All aircraft brighter than 11th magnitude, whether identifiable or UAPs, are detected. The multiple diagonal lines exhibit different lights on the aircraft and their horizontal width contains the real-time spectrum. The angular velocity per second is simply the angular length, in this case ~25 arcmin/s.*

## 5   DETECTION THRESHOLDS

We used an injection-and-recovery approach to measure the probability of detection of monochromatic emission in a single exposure as a function of the intensity of that emission. We established an empirical PSF by using the CIII emission line and the Hα emission line in Campbell's Star, which differed by ~10% in width. The 2D shape (spatial and spectral) offers an empirical proxy for the shape of a monochromatic emission line in our images, such as from a laser pulse. We cut out a region of 25 x 25 pixels, centred on the Hα emission line in Campbell's Star, and injected it into other images at random pixel locations. This randomly-placed proxy for monochromatic emission has a shape created by the PSF (optical and seeing) convolved with minor intrinsic size of the object and its spectroscopic width. The final FWHM is ~5 pixels (=6.5 arcsec), as expected from the known PSF and seeing. We scaled these mock injected monochromatic pulses to various total numbers of photons within the entire profile, from 170 to 620 photons. We added these PSF shapes to actual individual images, simulating a monochromatic pulse having a duration less than 1 sec. We placed the pulses at random locations within the image, both in between and coincident with stellar spectra.

We executed the difference-image analysis to determine if it "discovered" the synthetic pulses, running 100 trials for each pulse intensity. The fraction of injected pulses detected is shown graphically in Figure 14. We found the code successfully discovered 50% of the injected pulses that had at least 290 total photons in the profile. For reference, such a 1-sec pulse would appear to be magnitude 15.0, in a one-second exposure with a broadband filter. To clarify this threshold, in a hypothetical 100-sec exposure taken by any telescope this same 1-sec pulse would appear to be magnitude 20. None of the injected monochromatic pulses that contained fewer than 230 photons was detected. In contrast, 97% of the pulses having more than 400 photons were detected.

Thus, the nominal detection threshold at which 50% of the pulses would be detected is 290 photons total within the monochromatic pulse. *This 290-photon threshold represents the number of photons that must be detected in 1 sec such that half of such pulses would be detected.* We note that the threshold in our previous laser search was 650 photons. The improved threshold to 290 photons is due entirely to the Octopi focusing mechanism that decreased the PSF FWHM from ~5.5 to ~4.0 pixels. This threshold is an average over the full 2x3 deg field, but does not include the variation in PSF shape over the full field, thereby imposing a factor of ~2 variation in the true threshold.

For pulses lasting over 2 sec, the search algorithm has poorer sensitivity due to adjacent 1-sec exposures being "contaminated" by the continued emission, thereby diminishing the difference with the target image. Still, the onset of a pulse that lasts many seconds or minutes may still trigger the difference-image algorithm because the prior three exposures will have little or no flux, yielding a detectable difference. Continuous monochromatic emission would require ~2900 photons per sec for the 10% variations to reveal itself as a pseudo-pulse. A train of pulses of nanosecond duration and arriving $10^6$ per second would be detected here only as "continuous" monochromatic emission, requiring seeing variations for detection and yielding a detection threshold of $12^{th}$ mag.

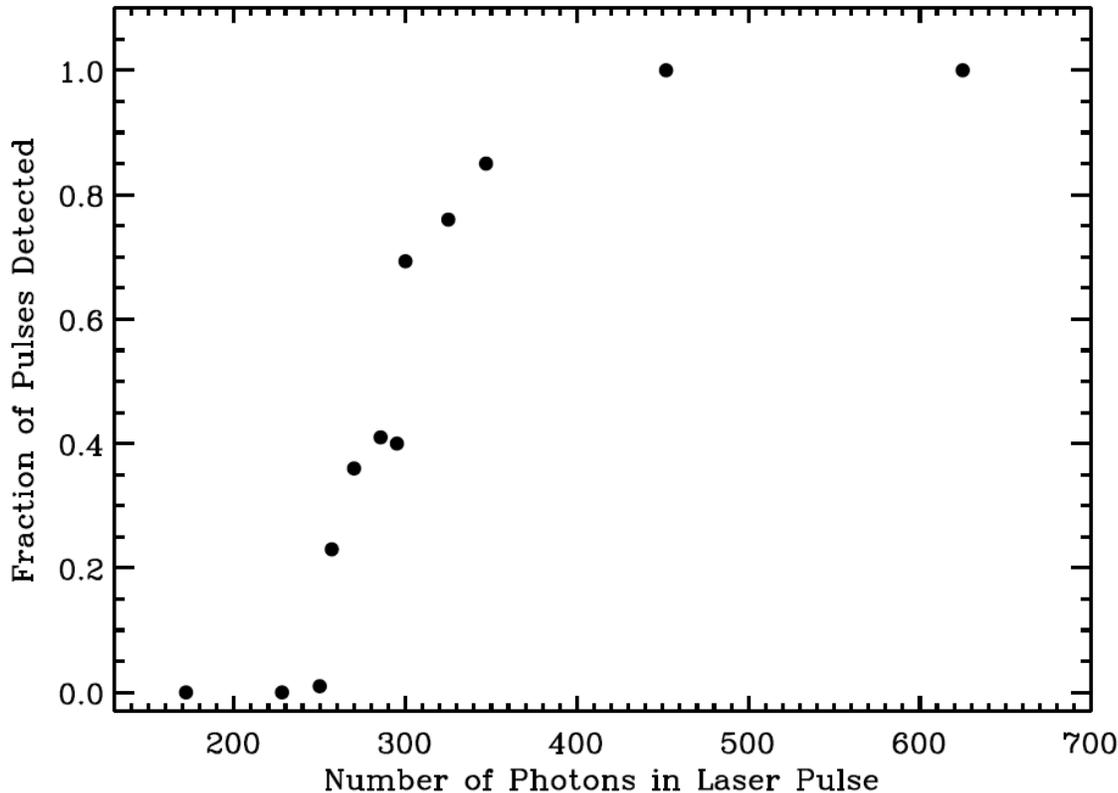

*Figure 14. Detectability vs. photons in a monochromatic pulse. The plot shows the fraction of injected monochromatic pulses that were detected blindly by the difference-image search algorithm as a function of the number of photons in the pulses. Pulses containing >450 photons are detected in 100% of the trials. Pulses containing 290 photons (total within the PSF) are detected in 50% of the trials, constituting a nominal detection threshold near magnitude, V = 15 mag*

The threshold determination described above pertains to monochromatic pulses that occur in empty sky, in between stellar spectra. For a monochromatic pulse that is spatially coincident with stellar spectrum, notably any pulse coming from within 1 arcsec of a star, the detectability is compromised by the competition with that background stellar light, depending on the star's brightness. Stars fainter than $15^{th}$ mag cause no degradation in detectability of pulses, while brighter stars require any pulses to deliver a fluence proportionately brighter to compete with the star. For example, a laser pulse coincident with the $12^{th}$ magnitude star must have a fluence 16x greater, i.e. ~4600 photons, to be detected. This degradation of detectability is consistent with the faintest emission-line stars we independently found, around 12th magnitude, as shown in Figures 9 and 10.

The detection thresholds quoted here pertain to the wavelength range, 500 to 600 nm, the traditional "V band". However, the QHY600 sensor, a Sony IMX455, has a response (quantum efficiency) that is approximately 10% better between 400 to 500 nm, improving slightly the detectability of pulses. At wavelengths less than 400 nm, the sensitivity falls rapidly to zero at 380 nm due to poor transmission of the optics, especially the lens group of the RASA telescope. For wavelengths longer than 600 nm the response falls nearly linearly to zero toward 950 nm, thus requiring proportionately higher fluence of photons for detection. In summary, searches for monochromatic pulses in the near-UV and near-IR require different optics and sensors.

# 6. HISTORICAL OPTICAL SETI

One wonders if past astronomical searches of the Galactic Plane would have detected laser emission having visible wavelengths 400 to 700 nm. Answering this question requires a study of all past observations of the Galactic Plane, which is beyond the scope of this paper. However, a hint stems from our blind rediscovery of Campbell's star (Section 4). The emission lines in Campbell's star, detected blindly here, consist of the spectrum of a planetary nebula that is so compact that the [OIII] forbidden lines are absent due to the high density of the gas. The spectrum also contains emission lines from the WC9 Wolf-Rayet central star, including rare emission lines from multiply-ionized nitrogen and carbon. Other Wolf-Rayet stars having rare emission lines, e.g., WC11, have also been detected (Margon et al. 2020, 2023). Thus, laser emission at arbitrary wavelengths coming from stars would have similarly been detected and pursued with follow-up spectroscopy. None has been reported. In Campbell's star, the combination of the two types of spectra, each of them rare individually, comprises a hybrid that requires careful examination to assess if the spectrum contains non-astrophysical spectral lines. Any such hybrid, especially if combined with fast winds or magnetically powered ionized gas, could masquerade as a possible non-astrophysical source, motivating further observations. This suggests that past surveys could have detected unexpected emission lines, including from technological sources of arbitrary wavelengths.

Remarkably, Campbell's star was discovered over 130 years ago by Williamina Fleming with an objective prism instrument similar to that employed here (Pickering 1891). She alerted Campbell (1893) to use visual spectroscopy at the Lick Observatory 36-inch telescope, revealing strong H$\alpha$ emission (Figure 10). Thus, historic surveys of the entire Milky Way Plane for emission line sources by objective prism photographic exposures and narrow-band imaging revealed emission lines at all optical wavelengths as faint as 12th magnitude. More recent spectroscopic and narrow-band surveys using CCD detectors surely constituted searches for laser emission, albeit unintentionally, providing fainter detection thresholds (see e.g., Margon et al. 2023).

Thus, a large domain of optical SETI parameter space was already searched by conventional spectroscopic and narrowband surveys. Indeed, one may wonder about the origin of the intuition, held by many astronomers and SETI researchers, that the detection of technological civilizations and their electromagnetic communication is unlikely in any given year, and may require decades or centuries, if ever. This intuition likely stems from past conventional astrophysics searches that revealed unexpected exotica, such as quasars, pulsars, and gamma ray bursts, but not technology (as far as we know).

Indeed, unintentional optical SETI began over a century ago. Objective-prism telescopes in the early 1900's revealed thousands of objects that exhibited spectral lines in emission, such as planetary nebulae, HII regions, T Tauri stars, Be stars, Wolf-Rayet stars, M dwarf flare stars, novae (cataclysmic variables), and active galactic nuclei, e.g., Fleming et al. (1907), Pickering (1912), Cannon & Pickering (1922). The detections included emission lines at wavelengths not anticipated or understood at the time, such as the "nebulium lines" due to forbidden transitions in doubly ionized oxygen [OIII]. The 225 000 stellar spectra in the original Henry Draper Catalog were examined individually by the vigilant Harvard "Computers", and these women discovered hundreds of peculiar spectra (e.g., Fleming et al. 1907, Fleming & Pickering 1908, 1910, Cannon & Pickering 1922).

Two such studies are worthy of special note. Fleming (1912) classified hundreds of stars with spectra so peculiar that the emission lines could not be identified at the time. These could have been due to technological sources. It is vital to realize that Fleming did not ignore those unexplainable spectra, nor did she deem them "noise", mistakes, or "photographic plate flaws", but instead she realized they were real and published them (Fleming 1912). Subsequent spectroscopy and analysis showed they were Wolf-Rayet stars, planetary nebulae, and flare stars, etc., all unexpected at the time, and thus brilliantly discovered. But they were not extraterrestrial technology, thus constituting an unannounced SETI non-detection.

Equally remarkable is that Maury (1897) examined high resolution spectra of over 4000 stars, noting unidentified absorption lines, strange patterns, and varying line widths. But she found no emission lines that later turned out to be technological. In summary, Maury, Fleming, Cannon, and Pickering, over 100 years ago, provided a major SETI non-detection and an associated upper limit of the occurrence of lasers. This profound non-detection was simply not recognized as such because SETI did not exist. Their upper limit for optical lasers can be calculated from the 225 000 stars that exhibited no laser emission. Their spectral resolution was roughly $\lambda/\Delta\lambda \sim 100$, capable of detecting narrow laser beams directed toward Earth. Adopting a benchmark light-beam opening angle of 1 arcsec, the upper limits of laser power are $\sim 2 \times 10^{-14}$ of each star's luminosity.

By 1991, spectra were obtained and inspected for over 359 000 stars, characterized in the extended HD catalog (Cannon & Mayall 1949), the Gliese Catalog of Nearby Stars (Gliese & Jahreiß 1979), and the Yale Bright Star Catalog (Hoffleit & Jaschek 1991). Stars with detected emission lines were noted, and among them unidentified and non-astrophysical transitions would have been detected and noted. None were so noted.

One may wonder if laser emission in a spectrum might have been typically overlooked or ignored as some flaw. Such mistakes were rare for two reasons. First, it was the general practice to record all empirical data and phenomena for the star catalogs, especially emission lines. Indeed, most emission lines noted in the extended Henry Draper Catalog and Gliese Catalog were based on just a single spectrum as obtaining multiple spectra of so many stars was not practical. Further, emission lines were known to be variable, such as from flare stars, novae, and T Tauri stars (Joy 1949, Marcy 1980), motivating mention of the emission in the catalog even if seen only once. Second, emission lines in photographic spectra were distinguished from plate flaws or cosmic rays by their location precisely coincident with a stellar spectrum. Indeed, any real emission line must exhibit the full width of the stellar spectrum (perpendicular to dispersion) and an extent along the length of the spectrum consistent with the spectral resolution. Detecting an emission line right on a spectrum is easier than detecting randomly located point sources in a star-field image (e.g., Villarroel et al. 2021). Each spectrum dictates exactly where to look.

Thus, the spectra of over 359 000 stars could have revealed emission at unusual wavelengths if the intensity was above the detection threshold. Laser pulses arriving frequently would have a higher probability of a fortuitously timed spectrum than those arriving infrequently. Further, spectroscopy of thousands of stars obtained since 1980, recorded with large telescopes, modern spectrometers and CCD detectors would also have exhibited any laser emission, such as the spectra of 5600 stars taken at the Keck Observatory 10-meter telescope (Tellis & Marcy 1917).

In addition, many searches for broadband optical pulses of sub-second duration have been performed, covering over half the sky (e.g. Wright et al. 2001, Stone et al. 2005, Howard et al. 2004, Howard et al. 2007, Maire et al. 2022). *No optical pulses were detected.* However, Villarroel et al. 2020, 2021, 2022a, 2022b, and Solano et al. (2023) have discovered simultaneous, multiple star-like transients, often in alignment, that appear in single snapshots from the Palomar Sky Survey taken in the 1950's. In one case, nine point sources appeared within half an hour within 10 arcmin on a photographic plate taken in 1950, but the sources are absent on previous and later image (Villarroel et al. 2021). On another Palomar photographic plate taken in 1952, three point-like objects, ~16$^{th}$ mag within 10 arcsec, do not appear in later images. These detections remain mysteries.

## 7. DISCUSSION

We carried out a wide-field optical spectroscopic survey for monochromatic emission, with a time-resolution of 1 s. The region of the sky surveyed here consisted of two swaths that were ~2 deg wide and located 2 deg above and 2 deg below the Galactic Plane, between Galactic longitude -4 and 249 deg. This survey adds to the previously observed swath 2 degrees wide centered directly on the Galactic Plane, constituting a 6-degree wide swath. No laser emission, pulsed or continuous, was found, with a detection threshold of V~15.0 mag during 1 s for pulses and 12.5 mag for continuous sources.

Among the 36 candidates found here, nearly all were explainable as astrophysical sources whose flux at some wavelength appeared to vary by 5 to 10% due to momentary variations in seeing. Some sources were known astrophysical emission-line objects, including planetary nebulae, Ae/Be stars, Wolf-Rayet stars, and dMe magnetic flare stars. Some candidates were merely stars whose photospheric continuum flux varied due to seeing, especially at wavelengths where molecular opacities caused high flux naturally. The faintest of these constant-flux astrophysical sources had V = 12.0 mag, and most were V = 8 to 11 mag. These detections of known sources ensure that monochromatic sources brighter than 12th magnitude and constant in time during many seconds could be detected here, including from lasers. As such detections depend on ~10% seeing variations, monochromatic pulses lasting less than ~1 s and brighter than ~15$^{th}$ mag could also be detected, including from lasers. A few aircraft triggered the image-difference algorithm, and in one case mere Poisson noise triggered an apparent detection that was a fluctuation at the level of 4 to 5σ.

This program was designed to detect sub-second monochromatic pulses, by virtue of 1-sec exposure times and unambiguous monochromatic light that creates a well-sampled point spread function ~5 pixels across (PSF). Such monochromatic pulses would have produced a well-defined PSF shape on one image (or perhaps on two successive images), for which the difference-image algorithm was optimized. In contrast, cosmic rays, local radioactivity, radioactive fall-out, or gamma rays would deposit electrons in only one or a few adjacent pixels, which were easily distinguished from the smooth 5-pixel wide profile, in both dimensions, of true monochromatic pulses. Conventional astrophysical surveys that employ longer exposures of 30 seconds to 30 minutes suffer from dilution of the source by background sky light, diminishing the detectability of a sub-second pulses. Here, monochromatic emission lasting many minutes or indefinitely would also trigger the image-difference search algorithm due to seeing changes, but with sensitivity poorer by a factor of ~10 compared to that of a sub-second pulse. No such monochromatic light was found, neither pulsed nor persistent.

The survey has shortcomings. The observations of each field were limited to 600 exposures of 1 s each. Any monochromatic source directed toward Earth that happened to be "off" during those 10 minutes would obviously not be detected. Detectability of pulses declines with increasing time interval between pulses for any cadences slower than one per 10 minutes. Conversely, for pulse intervals shorter than 1 sec, the pulse train is unresolved in time, mimicking a constant source. In that case, the detection threshold is ~10x higher (worse), ~11th mag, as detection depends on seeing variations. Also, the survey was intended to be uniform. But the images suffer from seeing variations from hour to hour and also PSF variations across the field of view. In addition, the observations reported here benefited from a novel tip-tilt focusing device, the Octopi, that improved the PSF width by ~30% relative to observations within 2 deg of the Galactic mid-plane reported in Marcy & Tellis (2023).

The threshold of 290 photons for detection (with 50% probability) of a single pulse translates into a threshold fluence per unit area of 14 500 photons per square meter at the Earth's surface for monochromatic pulses of duration less than 1 sec. Here we assumed a 0.278-m RASA telescope, its efficiency between 450 – 800 nm, and the obstruction by the camera and lens group at prime focus (see Marcy, Tellis, and Wishnow 2021, 2022). For a pulse lasting 1 sec, a fluence of 14 500 photons $m^{-2}$ corresponds to V = 15.0 mag. For wavelengths below 450 nm and above 800 nm the quantum efficiency drops below 50% of peak QE (at ~600 nm), thus requiring a greater intensity for detection. Atmospheric extinction raises this threshold fluence at the top of the Earth's atmosphere by a few percent.

One may calculate the required laser power to produce a fluence of our detection threshold. A benchmark laser located 100 ly from Earth that is diffraction-limited with a 10-meter diameter aperture emits a beam with an opening angle of ~0.01 arcsec at a wavelength of ~500 nm. A photon fluence at Earth of 14,500 photons per square meter requires a benchmark laser to output a power of 45 MW during a 1 sec pulse. For a diffraction-limited benchmark laser located 1000 ly away, a laser power of 4.5 GW is required. Extinction from interstellar dust will increase this requirement by ~30%, requiring 6 GW. The laser beam footprint at Earth would have a diameter of 0.3 to 3 AU, respectively, a fraction of the area of the inner Solar System. Such footprints, launched 100 or 1000 ly away, suggest a laser beam directed by luck or purpose toward the Solar System. A smaller transmitter aperture requires greater laser power for it to be detectable by a given instrument. The laser power required increases as the inverse square of the aperture's diameter.

We note that a monochromatic pulse lasting only a few nanoseconds, microseconds, or milliseconds would have been detected here as long as the fluence exceeds 14500 photons $m^{-2}$, i.e. 15th mag during 1 sec. Past surveys would have been unlikely to catch such fleeting signals, especially those having low cadence that makes confirmation difficult. Here, remarkably short pulses are detectable, if bright enough. However, pulses as short as nanoseconds or microseconds may arrive with high pulse rates, greater than 1 Hz, thereby appearing here as continuous emission.

## 8. SUMMARY

We searched for optical laser pulses coming from a 6 degree-wide belt along the plane of our Milky Way Galaxy. We would have detected laser pulses as faint as 15th magnitude, i.e. 14 500 photons per square meter hitting the Earth's surface, if the pulses arrived every few seconds, few minutes, or hour. For continuous laser emission, the detection threshold is 10x higher. We found no evidence of optical lasers, pulsed or continuous, within optical wavelengths 380 to 950 nm.

Non-detections offer value in two ways. They downgrade theories that otherwise would remain plausible, notably that interstellar laser communication might be common in the Milky Way Galaxy (Schwartz & Townes 1961), emitted by civilizations and by their plausible billions of robotic spacecraft (e.g. Freitas 1980, Gillon 2014, Hippke 2020, 2021ab; Gertz 2018, 2021, Gillum 2022, Gertz & Marcy 2022). The non-detections here indicate that either the Milky Way is not filled with laser beams, or that advanced technologies purposely hide their beams from Earth (Forgan 2014). Non-detections also specify the domain that is less fertile for future searches, specifically pulses of optical light delivering more than ~20 000 photons per square meter.

One may imagine this search to be folly because laser beams are "narrow" and the Galaxy is "enormous", implying *a priori* beams would rarely intersect Earth. However, the density of civilizations and their laser transmitters remains unknown. The prospects remain open for our Galaxy to contain thousands of civilizations, consistent with the billions of Earthsize, luke-warm planets in the Galaxy (Petigura et al. 2013, Dressing & Charbonneau 2013). A typical civilization may launch billions of robotic laser-communication nodes during each millenium, including repeater-stations along a linear conduit (Gertz & Marcy 2022). The beams could have any opening angle, multiple wavelengths, and a variety of pulse cadences, yielding an arbitrarily high density of laser beams. Thus, the density of laser beams in the Galaxy was unconstrained both empirically and theoretically, no matter our intuition.

Past spectroscopic surveys for optical lasers of more than 5000 stars also found none (Reines and Marcy 2002, Tellis and Marcy 2015, 2017, Marcy & Tellis 2023, Marcy 2021, Marcy et al. 2022, Tellis 2022 private communication, Zuckerman et al. 2023). Stars of all spectral types were searched including O, B, A, F, G, K, and M-type stars, and some giants and supergiants. No laser emission was found, nor even plausible candidates. Searches for broadband optical and IR pulses have also yielded no definitive detections of laser pulses (cf., Howard et al. 2007, Maire et al. 2019, Villarroel et al. 2023, Acharyya et al. 2023, Korzoun et al. 2023).

Impressively, technological emission lines were also not discovered by any all-sky surveys, including the 359 000 stars spectroscopically scrutinized and described in the Extended HD catalog (Cannon & Mayall 1949), the Gliese Catalog of Nearby Stars (Gliese & JahreiS 1979), and the Yale Bright Star Catalog (Hoffleit & Jaschek 1991). Further, during the past century, hundreds of all-sky optical and radio-wave surveys have revealed many objects exhibiting unexpected and extraordinary spectral emission such as planetary nebulae (including forbidden lines), flare stars, compact HII regions, Wolf-Rayet stars, supernova remnants, quasars, high redshift "Lyman-break" galaxies, and optical gamma-ray bursts. Those surveys discovered tens of thousands of unexpected astrophysical exotica. But no lasers were found nor narrowband radio emission.

Many deep SETI surveys have been performed. Cherenkov telescopes of 12-meter aperture have explicitly searched for optical SETI pulses (Acharyya et al. 2023, Korzoun et al. 2023) achieving extraordinarily low thresholds of ~10 photons $m^{-2}$, but finding no optical pulses. Hundreds of broadband radio-wave surveys of the entire sky revealed many radio sources, including over 3000 pulsars and 1 million quasars, but none turned out to be narrowband technological radio-wave signals. Indeed, over 300 000 stars were surveyed in explicit radio SETI surveys, yielding only non-detections (Garrett & Siemion 2022; Wlodarczyk-Sroka, Garrett, Siemion 2021).

In summary, the non-detections described here of laser beams within our Milky Way Galaxy add to 64 years of non-detections by past optical and radio SETI observations. In addition, thousands of all-sky surveys, performed at all wavelengths, discovered many extraordinary astrophysical objects, but revealed no signs of technology. Hypotheses that presume the Milky Way Galaxy is teeming with technological beings must be demoted, a profound realization. The dearth may be due to the rare evolutionary ascent of sentient beings or to their self-destruction (e.g. Bostrom 2014, Ord 2020, Garrett 2024).

## ACKNOWLEDGEMENTS

This work benefitted from valuable communications with Beatriz Villarroel, Brian Hill, John Gertz, Ed Wishnow, Franklin Antonio, Ben Zuckerman, Susan Kegley, Paul Horowitz, and Dan Werthimer. We thank the team at Space Laser Awareness for outstanding support. We thank the anonymous referee and the editors at MNRAS for valuable suggestions that greatly improved the manuscript.

## DATA AVAILABILITY

This paper is based on raw CMOS sensor images obtained with Space Laser Awareness double objective prism telescopes. The 148000 raw images are 125 MB each, totaling 19 TB. They are located on two peripheral disks that are not online. All images are available to the public upon the request of G.M., and a transfer method must be identified.

This paper was typeset from Microsoft WORD document prepared by the author.